# Towards Solving Cocktail-Party: The First Method to Build a Realistic Dataset with Ground Truths for Speech Separation

**Rawad MELHEM** *
*Higher Institute for Applied Sciences and Technology, Damascus, Syria, rawad.melhem@hiast.edu.sy*

**Assef JAFAR**
*Higher Institute for Applied Sciences and Technology, Damascus, Syria, assef.jafar@hiast.edu.sy*

**Oumayma AL DAKKAK**
*Higher Institute for Applied Sciences and Technology, Damascus, Syria, oumayma.dakkak@hiast.edu.sy*

* Author to whom correspondence should be addressed

*Abstract:* - A significant improvement occurred in recent years towards the solution of the cocktail-party problem. In fact, much attention has been drawn to supervised learning methods using synthetic mixtures datasets despite their being not representative of real-world mixtures. The difficulty in building a realistic dataset led researchers to use unsupervised-learning based methods, because of their ability to handle realistic mixtures directly. The results of unsupervised methods are still unconvincing. In this paper, a method is introduced to create a realistic dataset with ground truth sources for speech separation. The main problem in designing a realistic dataset is the unavailability of ground truths for speakers' signals, so a method is suggested to record two speakers simultaneously and obtain the ground truth for each speaker. The proposed method utilizes a MATLAB function which exploits a full duplex sound card to record and playback audio files at the same time. TIMIT corpus was used to implement the proposed method, and design Realistic_TIMIT_2mix dataset. Evaluation is carried out on three datasets, and experiments show that the proposed dataset improved SI-SDR by more than 1.5 dB and PESQ by 0.5 approximately. The performance of the proposed method was evaluated on different distances between the microphone and the speakers, the results showed that the proposed method made the learned model more stable when the distance changes.

## 1. INTRODUCTION

Cocktail Party problem was first raised more than half a century ago [1]; the required solution is to separate the individual speech signals of different speakers from a mixture of multiple speakers using a single channel in a realistic environment. During the last years, several deep learning models were proposed to improve the accuracy of separation, and a great improvement has been achieved. First, supervised learning methods have been widely used to propose solutions, The authors in [2] optimized embeddings with BLSTM layers for efficient clustering and a 6dB improvement in speech separation. [3] introduced a deep learning framework using attractor points in a high-dimensional embedding space for single-channel speech separation. SepFormer, an RNN-free Transformer-based network in [4], achieved state-of-the-art results with SI-SDR of 22.3 dB and 19.5 dB on WSJ0-2/3mix datasets. In [5], a joint feature map combining convolution encoded features and spectrogram improved embedding and clustering by considering both time and frequency domains. [6, 7] proposed novel deep learning models for multi-talker speech separation, addressing the cocktail-party problem and supporting permutation invariant training. [8] presented TF-GridNet, leveraging multi-path blocks with self-attention modules for monaural talker-independent speaker separation. [9] introduced TasNet, a waveform-based audio separation network that estimates source masks on encoder outputs, simplifying the task without frequency decomposition. Conv-TasNet is introduced in [10] as a fully convolutional time-domain audio separation network for end-to-end speech separation. It utilizes a linear encoder to optimize speaker separation by applying masks to the encoder output, inverted by the decoder. In [11], DPRNN is presented for modeling long sequences. Using a deep structure, DPRNN splits and processes the input iteratively, enabling effective modeling of extremely long sequences. DPTNet is introduced in [12] for end-to-end speech separation, enhancing context-awareness in speech sequence modeling. The improved transformer incorporates a recurrent neural network to capture order information without positional encodings. Wavesplit, introduced in [13], is an end-to-end source separation system that infers source representations and estimates source signals, addressing the permutation problem. Tiny-Sepformer, a compact Transformer network for speech separation, is presented in [14], achieving comparable performance to vanilla Sepformer with reduced model size using Convolution-Attention blocks and parameter sharing techniques. All the previous papers used clean synthetic dataset WSJ0_2mix [2]. This dataset contains clean, synthetic, read speech and near-field utterances. Researchers obtained very good results with the previous methods; where SI-SDR reached 23.4 dB in [8], and PESQ reached 3.51 in [15]. However, these results were obtained in the same conditions in which the models were trained on (ideal clean instantaneous mixing). Unfortunately, the performance decreases in the realistic experiments [16].

Second, some researchers tried to make the training dataset more similar to the reality. As in deep learning, the more similar the training data to reality is, the more accurate the learned model is; noisy synthetic mixtures were made, where the noise was either added mathematically to the clean speaker signal like WHAM! [17], WHAMR! [18], and LibriMix [19], or was recorded with the speaker signal to be more realistic like CHiME-3 [20], CHiME-5 [21], Mixer6 [22], and VoxCeleb [23], then the noisy synthetic mixtures would be created by adding the noisy recorded speakers signals together mathematically (sample to sample). In [24], the authors show through experiments that training on noisy oracle sources can lead to significant improvements in speech separation performance, particularly in noisy environments where traditional methods may not perform well. REAL-M is introduced in [16] as a realistic dataset for speech separation, utterances are collected by asking contributors to read predefined sentences from LibriSpeech [25] dataset simultaneously in different acoustic environments using different recording devices to reflect the real-world scenarios. REAL-M is a real-life dataset but without ground truths, therefore it can only be used in unsupervised learning methods.

Indeed, adding two speakers' signals mathematically differs from real-world mixture, the main drawback in making a realistic dataset for speech separation is finding the ground truth for each speaker, since it is impossible to record the same sound from the same speaker twice, they will certainly differ in amplitude, frequency, and duration. That is why recording two speakers simultaneously then recording each speaker alone will not give an accurate ground truth for training.

The unavailability of the realistic datasets spurred the researchers to think of unsupervised learning methods, by dealing with the realistic mixtures directly. Yannan Wang et al. propose an unsupervised single-channel speech separation approach in [26] using a deep neural network to predict the gender of the speakers in the mixture and then separate the speech signals based on their predicted gender. The authors show that the use of gender information can improve the accuracy of speech separation. In [27], Kohei Saijo et al. introduce a new unsupervised speech separation algorithm that uses a cycle-consistent adversarial training approach to improve accuracy and stability; the used loss function leads to an explicit reduction in the distortions. The authors in [28] offer an unsupervised method for speech separation, MixIT which uses MoM as an input and separate them into variable number of latent sources, then remix them to make the original mixtures. This method suffers from an over-separation problem. Teacher-Student framework is used in [29] to address the over-separation problem, where teacher model is trained using mixture of mixtures and MixIT, then the separated sources produced by teacher model are used as pseudo-targets to train student model using PIT [7]. In [30], a new adaptation technique for unsupervised speech separation is proposed, that uses heterogeneous neural networks to produce high confidence pseudo labels of unlabeled real speech mixtures, then these labels are updated iteratively and used to refine the neural networks to produce more reliable pseudo labels for real mixture. This framework outperformed the previous unsupervised methods, but there remain some errors in pseudo labels that need to be removed in the future. Unsupervised speech separation is still a challenging open problem and requires hard efforts to improve the accuracy.

So far, all the training datasets for speech separation with or without noise are synthetic or instantaneous. This means the mixture signal is artificial and is made using the digital addition (mathematic sum) of signals. Synthetic mixture differs from the realistic one which is the main reason behind performance degradation of models trained on synthetic datasets. In this paper, a new method is introduced to build a realistic dataset with an acquisition of ground truth for each speaker; to deploy the method, TIMIT corpus was used to create

*Realistic_TIMIT_2mix*. The proposed algorithm depends on playing and recording an audio file simultaneously using MATLAB function *APR*. *APR* is applied for each speaker file alone to get ground truth for each speaker, then *APR* is applied for the two audio files, one on the left channel and the other on the right channel of the audio output device, and recording them (to get the mixture signal) using a microphone while playing the signals.

To evaluate Realistic_TIMIT_2mix, the following methodology is applied. First, a synthetic dataset Synthetic_TIMIT_2mix was created using the same files of Realistic_TIMIT_2mix where every realistic mixture is mapped into a synthetic mixture from the same both speakers. Second, two copies of a base model for speech separation were trained, one on Realistic_TIMIT_2mix and the other on Synthetic_TIMIT_2mix. Third, the results of both models were compared using SI-SDR and PESQ metrics. Fourth, the performance metrics of two models were measured on different distances between the microphone and the speakers. To the best of our knowledge, this is the first realistic training dataset with ground truths for single-channel speech separation.

The rest of the paper is organized as follows. Section 2 explains the technique used. The proposed method is introduced in section 3. Section 4 demonstrates the effectiveness of the proposed method through experiments. Section 5 presents the results and analysis. Finally, the paper is concluded in section 6

## 2. AUDIOPLAYERRECORDER

Mathworks corporation introduced an object in R2017a version of MATLAB software called *APR*, which uses a full duplex sound device to read and write audio samples simultaneously in real time. Figure 1 (from MATLAB help documentation) demonstrates the way APR controls sound device. It needs exclusively ASIO driver for sound device for windows operating system, since it provides low latency of recording. This driver bridges between MATLAB software and the sound card of the PC. In the proposed algorithm a full duplex sound card is required which has two buffers one for storing data coming from microphone and the other for storing data to be sent to the audio output device. The synchronization of rendering and recording audio files is considerable, so if the synchronization is lost, MATLAB returns the number of lost samples.

Two cases may occur, overrun and underrun.

Overrun case: it happens during the recording operation, when the input buffer of the sound card is full, in this case new samples of the input signal are dropped.

Underrun case: it happens during the playing of audio file when the output buffer is empty, because the MATLAB algorithm doesn't supply samples to be played, so the output signal in this case is silence.

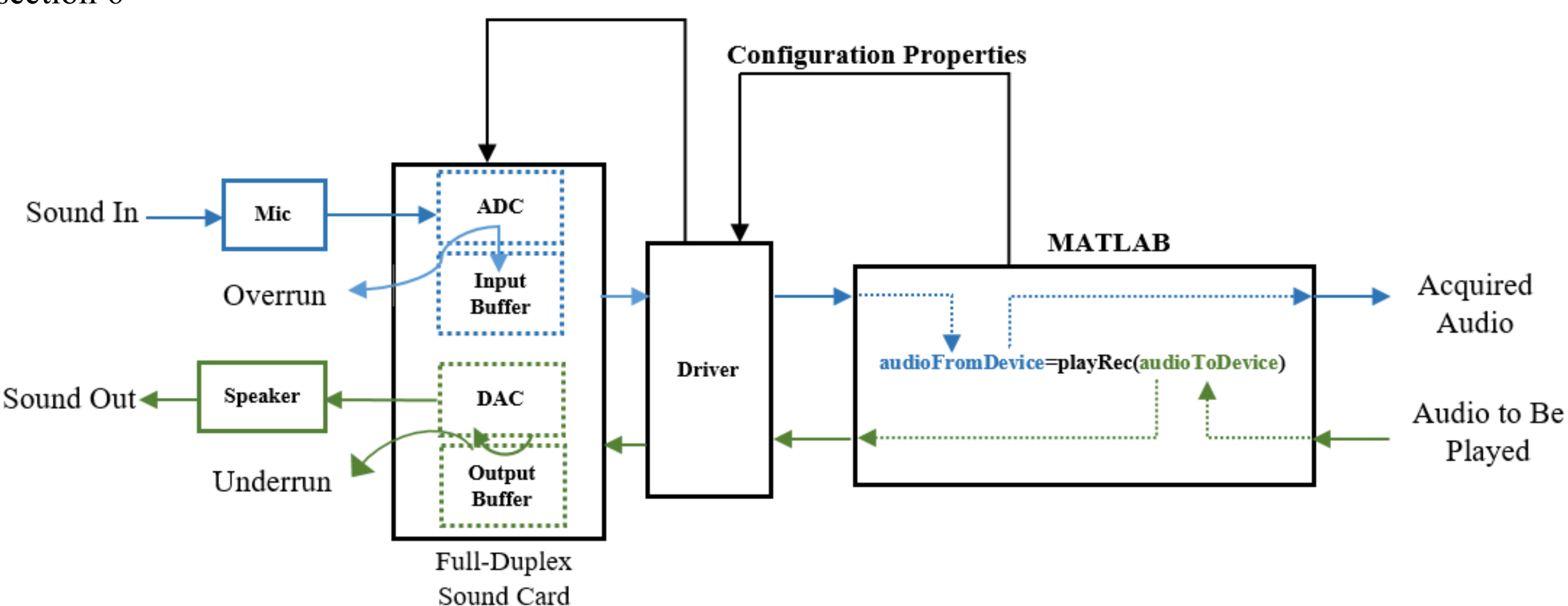

**Figure 1**. demonstrates how APR interacts with the full duplex sound card

These two cases will be avoided in the algorithm, APR returns the number of overrun and underrun lost samples, if they are bigger than zero the APR will be rerun.

Another kind of synchronization must be achieved, the synchronization between left and right channels. Playing two different audio files on dual channels must be at the same time without any delay, because any shift between the playing of the two files will make errors in the recording of the dataset. In our work, the sound card driver supports multichannel audio playback, so the audio data for each channel is stored in a separate section of the output buffer, and both channels are connected with DAC with a sampling rate of 192KHz. This sampling rate is very high compared to the sampling rate of audio files, and hence minimizes the time shift between left and right channels. To make sure that time shift is very small between both channels; it was measured by a *BK Precision 2566-MSO* Oscilloscope (with sampling rate of 2GS/s). It showed no delay between the channels.

## 3. PROPOSED METHOD

The synthetic mixture signal or the instantaneous mixing of *C* signals can be defined by equation 1.

$$X_s = \sum_{k=1}^{C} s_k \tag{1}$$

where $s_k$ is the k-th speaker signal. $C$ is the total number of speakers, and $X_s$ is the synthetic mixture signal. In real world the mixture signal $X_r$ is more complicated. It can be formulated as the result of application of nonlinear time-varying function $\mathcal{H}$ on the speakers' signals, as is written in equation 2.

$$X_r = \mathcal{H}(s_1, \dots, s_C) \tag{2}$$

$\mathcal{H}$ is usually approximated by the sum of convolution multiplication of each speech signal with the corresponding impulse response of the recording room [31]. However, in reality, it is much more complicated than that, because the impulse response itself is an approximation of the channel between the speaker and the microphone, and it is affected by many factors such as noise, wind, locations of speakers, nonlinearity of the microphone, and interference, etc.

In this paper, $X_r$ is directly recorded by a MATLAB function called *APR* that renders and records audio files simultaneously. The proposed algorithm depends on playing two audio files simultaneously, one on the left channel and the other on the right channel of the audio output device, and recording them at the same time to get the mixture signal. The synchronization issue is very important here, it must run the two audio files and record them at the same time. Any error in the synchronization will cause an error in the alignment of the played files and the corresponding recorded file (mixture signal), which leads to mistakes in the recording ground truths.

The proposed method was achieved in three steps: Audio Files Processing, Acquisition of Ground Truths, and Recording the Realistic Mixtures.

### 3.1. Audio files processing

The proposed method can be applied to any dataset; TIMIT corpus was chosen [32] for that. Each chosen audio file from TIMIT will be renamed by combining the dialect region, speaker ID, and sentence text, as is shown in figure 2. This new form of files names points to the path of each file in TIMIT folder and helps us to maximize the diversity of dialect and gender, and to make mixture names list.

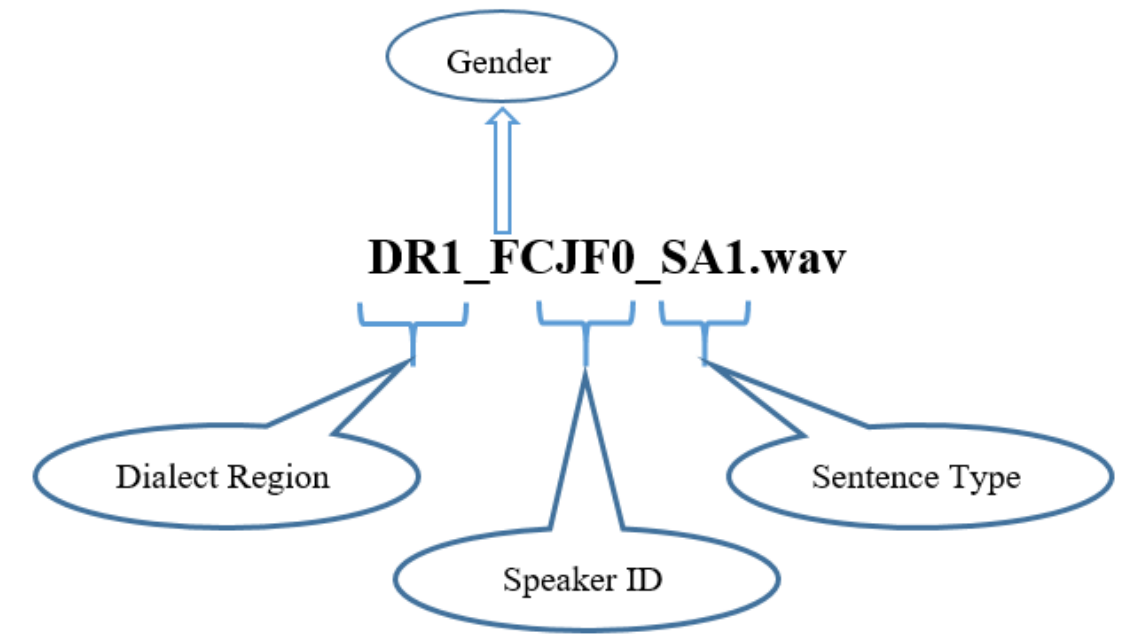

**Figure 2**. The new form of audio file names of TIMIT

TIMIT has 630 speakers from eight different dialects of American English. We constructed the set *S* of different wav files renamed as in figure 2, and also constructed mixture names list $L$ based on algorithm 1 given in [33]. This algorithm depends on four criteria: rejecting any mixture from two audio files by the same speaker, diversification of uttered speech in the files, maximizing the diversity of speakers within mixtures, and choosing files of similar lengths as much as possible to minimize padding.

### 3.2. Acquisition of the ground truths

The most important issue in this paper is the acquisition of ground truth for each speaker, because it is easy to record realistic mixtures but not the corresponding ground truths. It is not useful to consider the clean files of TIMIT as ground truths because the travelled audio signals to the input of the microphone differ from that coming out from the audio output device. The idea is to exploit *APR* function to render and record each speaker file alone with the same conditions used for the recorded mixture. In this case, the recorded signal could be considered as the ground truth for that speaker.

### 3.3. Recording the realistic mixture

After the acquisition of ground truth for each speaker, the realistic mixtures were recorded using *APR* function as described above.

In algorithm1 the instruction *APR(spk1)* means rendering and recording the wav file *spk1* using *APR* MATLAB object, and returning the recorded wav file, the lost samples number of overrun, and the lost samples number of underrun. In addition, the instruction *APR(spk1,spk2)* means rendering *spk1* on the left channel and *spk2* on the right channel and recording them simultaneously. It returns the realistic mixture wav file, the overrun lost samples number, and the underrun lost samples number.

**Algorithm 1** *Creating Realistic_TIMIT_2mix*
**Input:**

- $S$ – set of wav files from TIMIT train folder.
- $L$ – mixtures names list.

**Output:**

- $\boldsymbol{GTS}$ – Folder containing ground truths for speakers.
- $RealMix$ – Folder containing the realistic mixtures.

1: $i \leftarrow 0$
2: **while** $i < length(L)$ **do**
3: Extract wav files $(spk1, spk2)$ from $S$ correspond *L(i)*.
4: Down-sampling *spk1 & spk2* to 8KHz
5: $gts1, oRun, uRun = APR(spk1)$
/* *gts1* – ground truth for spk1.
*oRun* – overRun lost samples number.
*uRun* – underRun lost samples number.
*/
6: **if** $(oRun\ \boldsymbol{OR}\ uRun) > 0$ **then** ***go to*** 5
7: $store\ gts1$ in $GTS$
8: $gts2, oRun, uRun = APR(spk2)$
// *gts2* – ground truth for speaker2.
9: **if** $(oRun\ \boldsymbol{OR}\ uRun) > 0$ **then** ***go to*** 8
10: $store\ gts2$ in $GTS$
11: $rMix, oRun, uRun = APR(spk1\ \&\ spk2)$
// *rMix* – realistic Mixture.
12: **if** $(oRun\ \boldsymbol{OR}\ uRun) > 0$ **then** ***go to*** 11
13: $store\ rMix$ in $RealMix$
14: $i \leftarrow i + 1$
15: **end while**

# 4. IMPLEMENTATION

## 4.1. Constructing the dataset

Algorithm 1 implementation should be optimized to run in real time, in order to make the quality of *Realistic_TIMIT_2mix as* high as possible. Several factors could affect the quality of the dataset like hardware, operating system, code implementation, and system load. The algorithm1 was run on PC with high hardware specifications, with Windows 10, and the MATLAB code was optimized to be at minimum delay. All other unnecessary applications, processes, and services were stopped to minimize the system load.

Algorithms 1 was run on a PC having the following hardware: motherboard ROG STRIX Z390-F GAMING, CPU core i9, RAM 64GB, GPU GeForce RTX 2080 TI, and audio codec type ROG SupremeFX 8-Channel High-Definition Audio CODEC S1220A.

Recording *Realistic_TIMIT_2mix* was accomplished in a lab far away from noise and bubbles. The distance between the microphone and audio output device is about 2 meters. The right and left channels are 50 cm apart, see figure 3. *Realistic_TIMIT_2mix* comprises 30 h training set, 10 h validation set, and 5 h testing set.

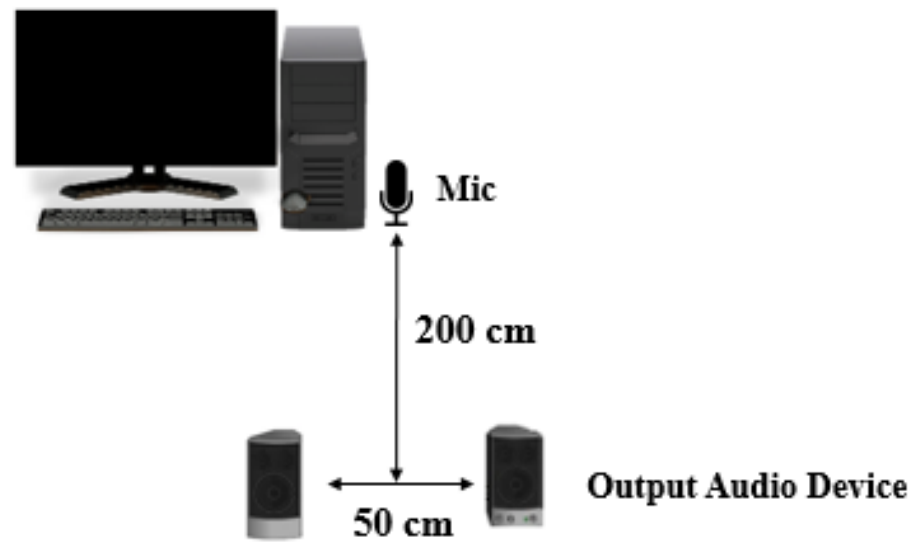

**Figure 3**. Experiment Setup of recording realistic dataset

## 4.2. Checking the validation of ground truths

There is a pressing issue that needs to be confirmed: Is the proposed realistic dataset valid for training or not? in other words: can we use the ground truths to train a model to predict a suitable mask for each speaker, so that the separated speech is as clean as possible?

Algorithm 2 is applied to check the validation of the ground truths of the speakers. This algorithm extracts the ideal mask for each speaker using the ground-truths, then reconstruct the estimated speaker using the extracted masks and the mixture. Three types of ideal masks could be used: IBM [34], IRM [35], and WFM [36] defined in equations 3, 4, and 5 respectively.

$$IBM_i(f,t) = \begin{cases} 1, & |S_i(f,t)| > |S_{j\neq i}(f,t)| \\ 0, & otherwise \end{cases} \quad (3)$$

$$IRM_i(f,t) = \frac{|S_i(f,t)|}{\sum_{j=1}^{C} |S_j(f,t)|} \quad (4)$$

$$WFM_i(f,t) = \frac{|S_i(f,t)|^2}{\sum_{j=1}^{C} |S_j(f,t)|^2} \quad (5)$$

where $S_i(f,t)$ is the complex-valued spectrograms of the ground truths, $i = 1, \dots, C$.

Algorithm 2 was applied for each mask type. Then the SI-SDR and PESQ metrics were calculated for the estimated speakers and the ground-truths.

**Algorithm 2** Validation of ground truths.
**Input:**

- $gts_i$ – ground truth for $spk_i$.
- $mix$ – mixture signal.

**Output:**

- $PESQ$ - Perceptual Evaluation Speech Quality.
- $SI_SDR$ - Scale Invariant Signal to *Distortion* Ratio

1: $S_i \leftarrow STFT(gts_i)$ // STFT-Short Time Fourier Transform
2: $M \leftarrow STFT(mix)$
3: $mask_i \leftarrow zeros(size(S_i))$
4: **for** each element $(f,t)$ in matrix $S_i$ **do**
5: $mask_i(f,t) \leftarrow \{IBM_i(f,t) or IRM_i(f,t) or\ WFT_i(f,t)\}$
6: **end for**
7: $est_{si} \leftarrow iSTFT\big((|M| \odot mask_i), \angle M\big)$
8: $SI_SDR \leftarrow si_sdr([gts_i],[est_{si}])$
9: $PESQ \leftarrow pesq([gts_i],[est_{si}])$
10: $\boldsymbol{return}\ SI_SDR, PESQ$

Algorithm 2 was repeated for every mixture in *Realistic_TIMIT_2mix.* Table 1 gives the averaged PESQ values and SI-SDR values.

**Table 1**. SI-SDR and PESQ average values for the ground truths evaluation.

| Metrics | *Realistic_TIMIT_2mix* | | |
|---|---|---|---|
| | IBM | IRM | **WFM** |
| SI-SDR (dB) | 14.1 | 13.9 | **14.3** |
| PESQ | 3.12 | 3.16 | **3.20** |

Table 1 shows that *Realistic_TIMIT_2mix* has high quality ground truths especially with WFM masks as compared with real conditions, where they got only SI-SDR values of 2.8 dB [16]. Hence, *Realistic_TIMIT_2mix* is suitable for training neural models.

## 4.3. Model configuration

In order to benchmark *Realistic_TIMIT_2mix*, and prove whether it is better than synthetic dataset or not, it is important to design synthetic mixtures from the same files used in building *Realistic_TIMIT_2mix,* the name given to it is *synthetic_TIMIT_2mix.*
For speech separation, the deep learning model given in [37] is used. It consists of 4 BGRU stacked layers followed by fully connected feed forward layer. The input features are the log spectral magnitude of the mixtures. The spectrum is obtained by STFT with 32ms window length, 8ms hop size, and hanning window. Figure 4 describes the model which maps each Time-Frequency bin into an embedding vector in a latent space. A clustering algorithm is then applied to estimate each mask.

The parameters of the base model are as follows. The optimizer algorithm is ADAM [38] with a training rate starting of $10^{-3}$, which will be halved if the validation error does not reduce in the last three epochs. The number of epochs is 300, and the batch size 128.

Two copies of the base model with the same parameters are used for training. One on *Realistic_TIMIT_2mix,* and the other on *synthetic_TIMIT_2mix*. The comparison between both results is given in section 5. For simplicity, the base model trained on *Realistic_TIMIT_2mix* is referred to as the **Realistic model**, and the model trained on *synthetic_TIMIT_2mix* the **Synthetic model**.

The validation loss for every model was drawn in figure 5. Realistic model loss reached its minimum (0.23) at epoch 157, and the Synthetic model loss reached its minimum (0.26) at epoch 143. The rapid convergence of the loss function observed in the synthetic model can be attributed to the high quality of the ground truth data, as the mixing process is considered ideal. Conversely, the loss function of the realistic model exhibited slower convergence due to the lower quality of the realistic dataset. In the realistic scenario, the audio signal traverses through the air from the audio output device to the microphone, causing slight distortions in the ground truth data. Overfitting in Synthetic is growing faster than overfitting in Realistic model. Early stop is used to handle overfitting.

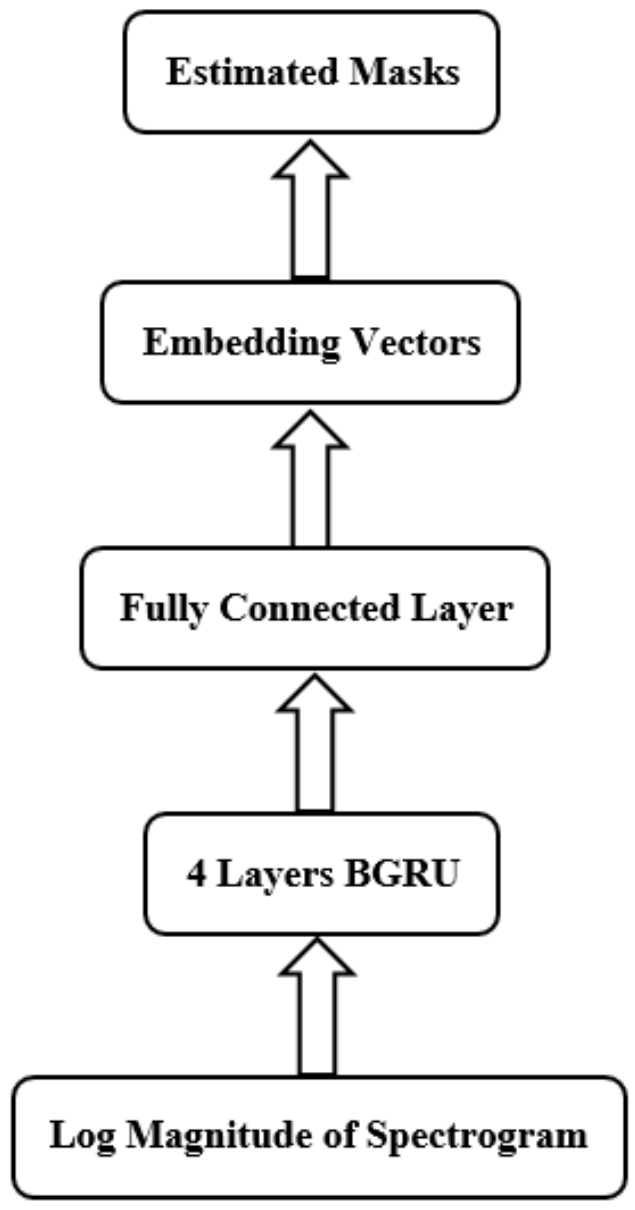

**Figure 4**. Base Model used for learning

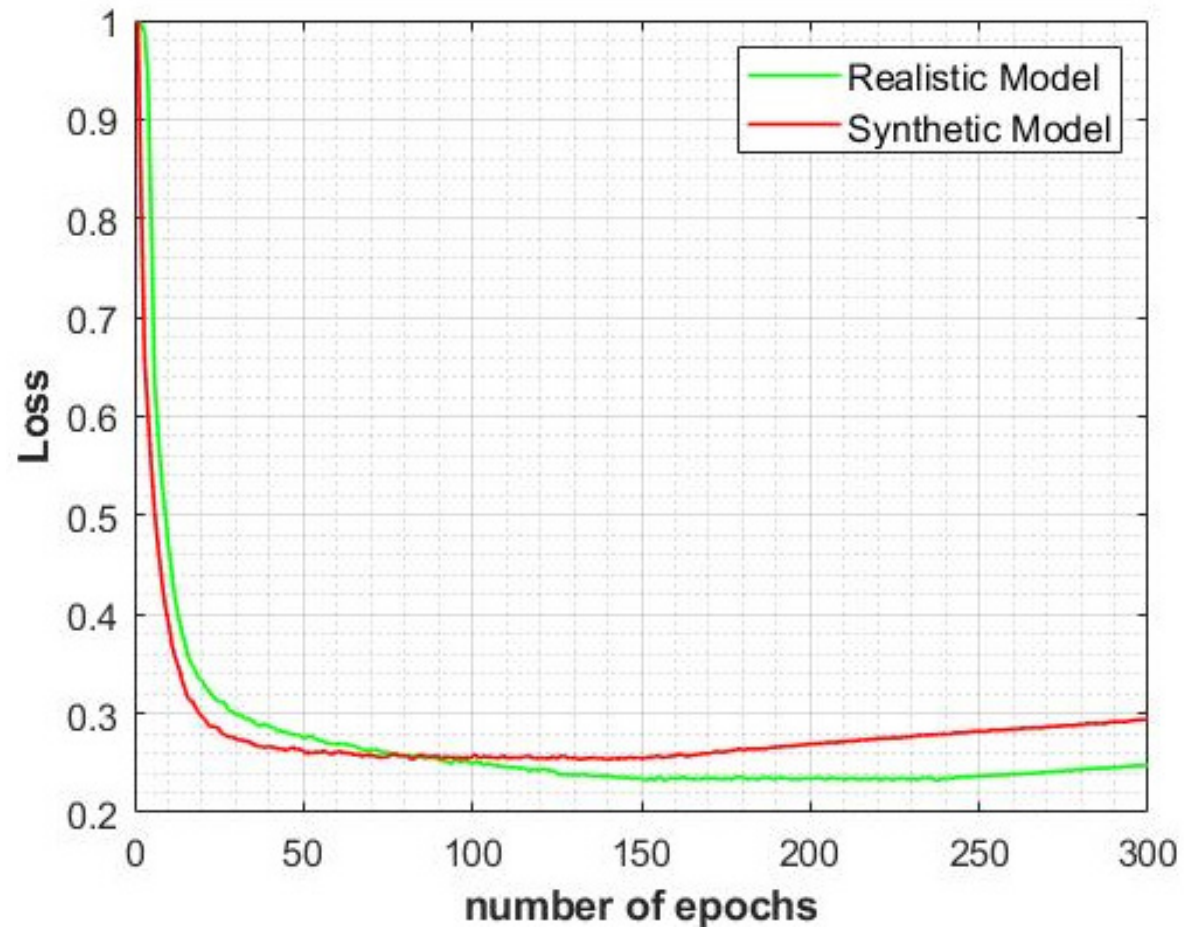

**Figure 5**. Loss Function for Synthetic and Realistic models

## 5. RESULTS AND ANALYSIS

Evaluation metrics used in this study are SI-SDR [39] to measure the speech separation accuracy, and PESQ to measure the quality of the reconstructed signals. Evaluation is achieved using three datasets as mentioned in the following section.

To benchmark the realistic model, it would be tested in three conditions: clean synthetic mixtures, noisy synthetic mixtures, and realistic mixtures, using the three testing datasets: Clean Libri2Mix, noisy Libri2Mix, and *Realistic Test*:

Libri2Mix: is a synthetic dataset which has two versions clean and noisy. Noisy Libri2Mix is based on LibriSpeech with ambient noise samples from WHAM! [17]

*Realistic Test*: the ultimate goal of speech separation is to separate mixtures in realistic conditions. That is why it is more meaningful to test on realistic mixtures than synthetic ones.

Five hours of *Realistic_TIMIT_2mix* were recorded for testing, using Test folder from TIMIT corpus (having other speakers than those used in the training). These 5 h mixtures were called *Realistic Test*.

In fact, the distance between the audio output device and the microphone was 2 meters when the training dataset '*Realistic_TIMIT_2mix'* was recorded. To check the performance on different distances, six copies of *Realistic Test* were designed with different distances between the audio output device and the microphone {0.5, 1, 1.5, 2, 2.5, 3} meters, the idea of changing the distance is to evaluate the robustness of the performance with the variation of the distance. In fact, distances exceeding 3 meters are uncommon in such settings, and they may introduce weaker signal strength and potential echo issues. Exploring these scenarios could be considered for future investigations.

On the other hand, it is well known that separation accuracy is very high in clean synthetic datasets (e.g. on WSJ0-2Mix) using the best neural model for separation, but it decreases sharply in synthetic mixtures with noise and reverberation (e.g. WHAMR!), and it is worse than that in real-life mixtures [16]. The separation results achieved by Realistic and Synthetic models on the three testing datasets are reported in tables 2, 3, and 4. It is clear that Realistic model outperformed the Synthetic model on all the three testing datasets.

**Table 2**. SI-SDR and PESQ on **Noisy** Libri2Mix using Realistic and Synthetic models.

| Model | SI-SDR(dB) | PESQ |
|---|---|---|
| Realistic Model | **11.08** | **2.01** |
| Synthetic Model | 9.81 | 1.54 |

**Table 3**. SI-SDR and PESQ on **Clean** Libri2Mix, using Realistic and Synthetic models.

| Model | SI-SDR(dB) | PESQ |
|---|---|---|
| Realistic Model | **13.65** | **2.68** |
| Synthetic Model | 12.40 | 2.19 |

**Table 4**. SI-SDR and PESQ on **Realistic Test** (2m distance between microphone and output device) using Realistic and Synthetic models.

| Model | SI-SDR(dB) | PESQ |
|---|---|---|
| Realistic Model | **8.66** | **2.09** |
| Synthetic Model | 7.01 | 1.68 |

Notably, Table 2 demonstrates the efficiency of the realistic model in handling noise during speech separation. It shows the outperformance of the Realistic model by about 1.27 dB. Since the model was trained using a realistic dataset, which inherently contains noise at even low levels, it learned to extract the speech signal from distorted mixtures. Additionally, Table 3 reveals that the realistic model achieved higher accuracy in separating clean mixtures compared to the synthetic model (by 1.25 dB). This could be attributed to the model's ability to extract speech structures in a distorted scenario, thereby enhancing its capability to perform separation in an ideal scenario. Furthermore, Table 4 shows an improvement in SI-SDR (exceeding 1.5 dB), since it presents the results obtained from realistic mixtures recorded using a microphone. The realistic model, having been trained on a dataset similar to these mixtures, demonstrates superior accuracy in separation.

In order to study the impact of changing the distance between microphone and audio output device, the performance of both models (realistic and synthetic) on different distances was measured. Figure 6 shows that the performance of realistic model is more stable and robust against distance changes; the SI-SDR using the synthetic model decreased to a value less than 2.5 dB when the distance increased to 3 meters, while the realistic model did not change a lot and it remained above 8 dB.

Figure 7 also reflects the stability of PESQ values, using the realistic model with the variation of distance, when compared with the use of the synthetic model. It shows the degradation of perceptual metric with the distance increase when using the synthetic model. Figures 6 and 7 provide evidence of the realistic model's stability when confronted with changes in distance. The model's proficiency in extracting speech structures from distorted mixtures allows it to perform better in separating sources at different distances. In contrast, the synthetic dataset assumes a zero distance between the output device and the microphone, as it is generated through mathematical summation. Previous results show the superiority of the realistic model and confirm the explicit effectiveness of the proposed method of recording in realistic acoustic conditions.

## 6. CONCLUSION

In conclusion, this paper tackled the cocktail-party problem in speech separation by highlighting the limitations of synthetic mixtures datasets in supervised learning. To address this, a method was proposed to create a realistic dataset with ground truth sources. By simultaneously recording two speakers using a full duplex sound card, ground truths were obtained. The Realistic_TIMIT_2mix dataset, constructed using the TIMIT corpus, outperformed other datasets in evaluation. The proposed dataset improved SI-SDR by over 1.5 dB and PESQ by approximately 0.5. The method's performance was also assessed under varying microphone-speaker distances, demonstrating increased stability. Overall, this research successfully introduced a realistic dataset for speech separation, resulting in enhanced performance and stability, thus contributing to the advancement of solving the cocktail-party problem.

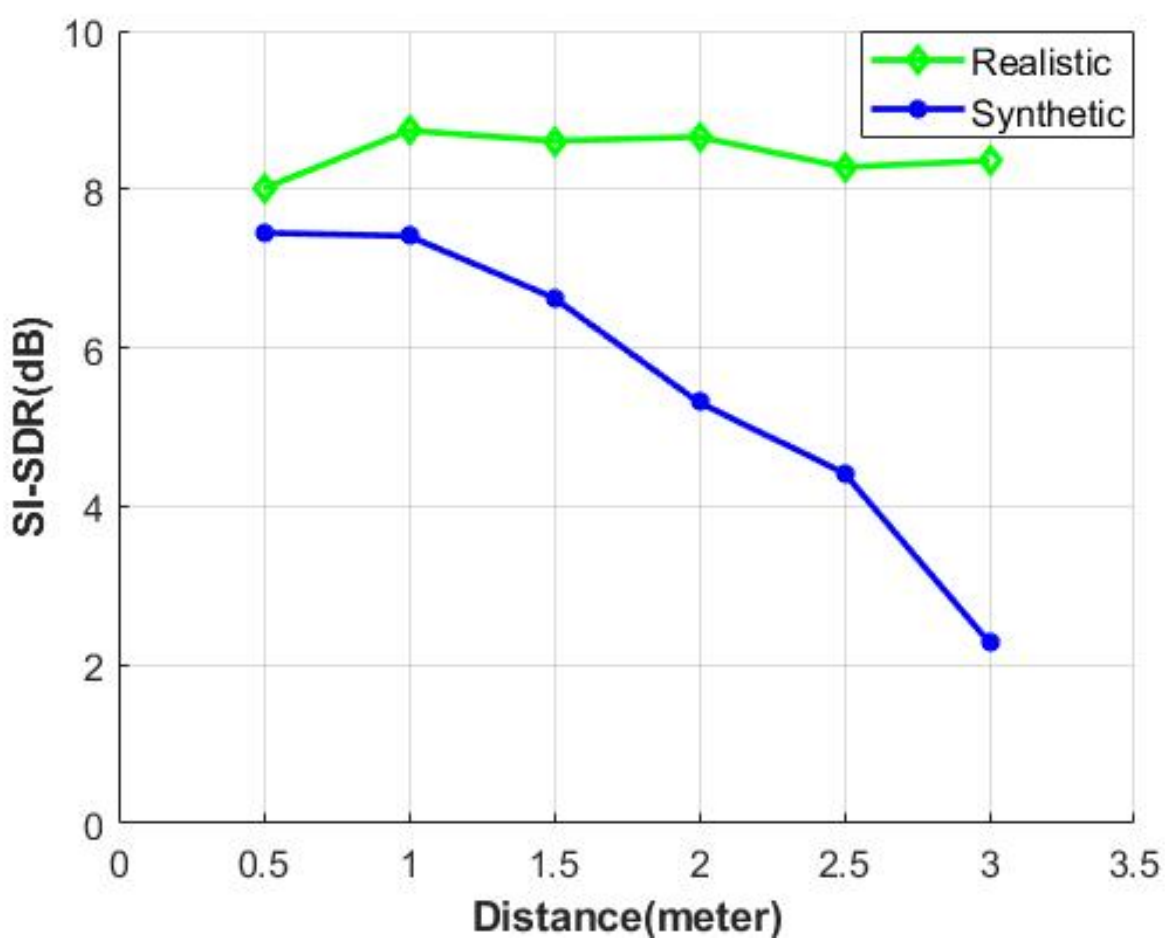

**Figure 6**. Effect of distance changes (between audio output device and mic) on SI-SDR for realistic and synthetic models.

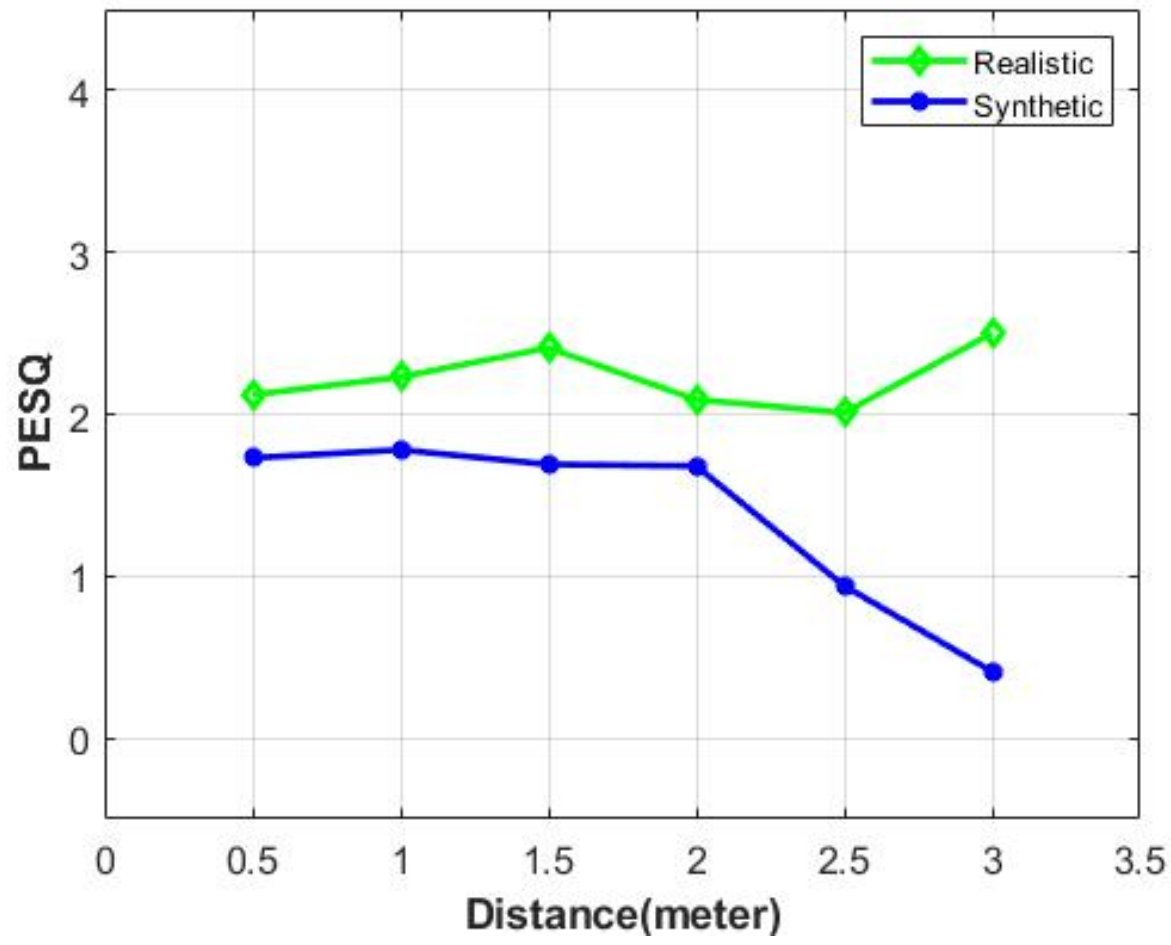

**Figure 7**. Effect of distance changes (between audio output device and mic) on PESQ values for realistic and synthetic models.

## ACKNOWLEDGMENTS

The authors thank the Department of Telecommunications at the Higher Institute for Applied Sciences and Technology for full support.

## ABBREVIATIONS

| Symbol | Description |
|---|---|
| APR | Audio Player Recorder |
| ASIO | Audio Stream Input/Output |
| BGRU | Bidirectional Gated Recurrent Unit Networks |
| BLSTM | Bidirectional Long-Short Term Memory |
| DAC | Digital to Analog Converter |
| DPRNN | Dual-Path Recurrent Neural Network |
| DPTNet | Dual-Path Transformer Network |
| IBM | Ideal Binary Mask |
| IRM | Ideal Ratio Mask |
| MixIT | Mixture Invariant Training |
| MoM | Mixture of mixtures |
| PESQ | Perceptual Evaluation of Speech Quality |
| PIT | Permutation invariant training |
| RNN | Recurrent Neural Network |
| SI-SDR | Scale Invariant Signal to Distortion Ratio |
| STFT | Short Time Fourier Transform |
| TIMIT | Texas Instruments/Massachusetts Institute of Technology |
| WFM | Wiener Filter-like Mask |

"